\documentclass[acmsmall,screen,nonacm]{acmart}

\usepackage{graphicx}
\usepackage{amsmath}
\usepackage{amsfonts}
\usepackage{amsthm}
\usepackage{braket}
\usepackage{listings}
\usepackage{subcaption}
\usepackage{xcolor}
\usepackage{wrapfig}

\theoremstyle{definition}
\newtheorem{definition}{Definition}
\newcommand{\symc}[1]{\llbracket#1\rrbracket}

\title{Unitary Expressions: A Necessary Abstraction for Extensible Quantum Programming Languages and Systems}

\author{Ed Younis}
\email{edyounis@lbl.gov}
\affiliation{%
    \institution{Lawrence Berkeley National Laboratory}
    \city{Berkeley}
    \state{California}
    \country{USA}
}

\ccsdesc[500]{Hardware~Quantum computation}
\ccsdesc[500]{Software and its engineering~Domain specific languages}

\keywords{Quantum Compilation, Equality Saturation, Quantum Simulation}

\begin{abstract}

Quantum gates are the fundamental instructions of digital quantum computers. Current programming languages, systems, and software development toolkits identify these operational gates by their titles, which requires a shared understanding of their meanings. However, in the continuously developing software ecosystem surrounding quantum computing—spanning high-level programming systems to low-level control stacks—this identification process is often error-prone, challenging to debug, maintenance-heavy, and resistant to change. In this paper, we propose replacing this nominal gate representation with a functional one. We introduce the OpenQudit system for describing, parsing, optimizing, analyzing, and utilizing programs comprising gates described as symbolic unitary expressions.

As part of this effort, we design the Qudit Gate Language (QGL), a unitary-specific expression language, and implement a differentiating just-in-time compiler in OpenQudit towards embedding this language in quantum programming languages and systems. Additionally, we have precisely designed and implemented the Qudit Virtual Machine (QVM) to evaluate quantum programs and their gradients efficiently. This evaluation is performed millions of times during the compilation of quantum programs. Our QVM can compute gradients approximately ten times faster than current leading numerical quantum compilation frameworks in the most common use cases. Altogether, the OpenQudit system is envisioned to (1) support many-level or qudit-based quantum systems, (2) enable the safe composition of program transformation tools, (3) accelerate circuit optimizers and transpilers, (4) enable compiler extensibility, and (5) provide a productive, simple-to-use interface to quantum practitioners.

\end{abstract}

\begin{document}

\maketitle

\section{Introduction}
\label{sec:introduction}

The digital circuit model first introduced by Feynman~\cite{feynman1986quantum} is the foundational programming model for quantum computing. This model sees quantum operations as gates applied to wires representing qubits. Given the rapid advancements in technology and science in this field, we believe that for the foreseeable future, every Quantum Information Science (QIS) practitioner will need to be an experimentalist, whether they focus on applications, infrastructure, or hardware development. Therefore, key criteria for compiler design include optimization potential, portability between quantum architectures and gate sets (transpilation), and overall extensibility and productivity. In particular, quantum compilers need to combine the quality of optimization requirements with flexible functionality to enable experimentation, leading to the following goals of quantum programming frameworks:
\begin{enumerate}
    \item \textbf{Expressibility}: the ability to represent all possible programs, including many-level systems such as those with qudit and many-qubit operations
    \item \textbf{Extensibility}: simplicity in adding new gates and associated optimizations
    \item \textbf{Accessibility}: tailored interfaces for practitioners to facilitate productive development
    \item \textbf{Safety}: support for safe composition and interoperability
    \item \textbf{Scalability}: the capacity to scale with the size of the system or program
\end{enumerate}
Existing quantum programming languages (e.g. OpenQASM~\cite{openqasm}, Q\#~\cite{qsharp}, QIR~\cite{QIRSpec2021}) and the associated infrastructures for program transformation (e.g. Qiskit~\cite{qiskit2024}, TKet~\cite{tket}, BQSKit~\cite{bqskit}, CIRQ~\cite{cirq}, cuQuantum~\cite{cuquantum}) meet some, but not all of these requirements.

Most importantly, they all use an opaque, nominal representation of gates for qubit-based operations, where the name of each gate reflects its functionality. While naming conventions work well for the few gates commonly discussed in the literature, they present tool extensibility and composability challenges. This can discourage experimentalists from exploring new, potentially impactful gates, as the surrounding software infrastructure often lags in support. This issue is especially pronounced in many-level systems, as adding qudit abstractions and optimization passes proves difficult. Moreover, composing tools and software stacks are typically required, if not strongly encouraged. However, tool composition can be risky and prone to errors due to discrepancies between gate functions. In general, maintaining an updated list of "well-understood" gate labels creates friction between different tools, hinders interoperability, leads to bugs, and ultimately slows down productivity.

Symbolic reasoning tools such as QUESO~\cite{queso} and Quartz~\cite{quartz} add a functional gate representation to enhance extensibility and discover new program transformations. Here, gate names can be associated with a symbolic mathematical representation. These tools synthesize symbolic rewrite rules over a gate set and utilize these rules for optimization. They can also internally check for circuit equality and thus have verification capability. Their scalability is currently limited to 3-4 qubit operations. The programming API is Java and C++, with extensions required to be programmed directly, which limits productivity. Extending these tools to support qudit systems, while appearing to be supported in theory, is not demonstrated in practice and, based on our experience, is likely to be challenging and require significant re-engineering. Integrating symbolic tools with other compilation pipelines currently requires first discovering a fixed set of rules, which is subsequently manually integrated into the other infrastructure.

Circuit synthesis-based compilers~\cite{bqskit, rakyta2022approaching, cincio2021machine} also use functional, unitary-based gate representations. They support functionality similar to symbolic tools but take a different implementation approach. Gates are represented as multivariate, unitary-valued functions, and circuits are the resulting function from composing all of its gates. Using this parameterized model, standard numerical methods like gradient descent are employed to identify a concrete circuit that best fits a given cost function. This process is commonly referred to as instantiation, helping to distinguish it from quantum program optimization, which aims to find an alternative program with the same semantics that uses fewer resources. Equality in these numerical tools is conducted over the instantiated circuits using quantum process or state distance measures.

Essentially, numerical approaches dynamically reveal circuit rewrite rules without needing predefined rules as input, leading to exceptional optimization potential. Additionally, these tools have successfully demonstrated qutrit and ququart (3- and 4-level qudits) program transformations~\cite{seifert2023exploring,younis2024qsweep}. Although one can quickly extend these tools to include new gates, this often requires providing analytical gradients for each new gate. This requirement can be tedious or even impossible for many users who lack the necessary expertise. Furthermore, the numeric nature of the tools requires numerous gradient evaluations. BQSKit is one such tool that overcomes this by partitioning a large program into small subcircuits and performing compilation passes on the subcircuits. While this method enables scalability, these compilers still demand highly optimized gate unitary and gradient evaluations, further adding to the extension complexity.



\begin{figure}[t]
    \centering
    \includegraphics[trim={6.5cm 0 0 0},clip,width=\linewidth]{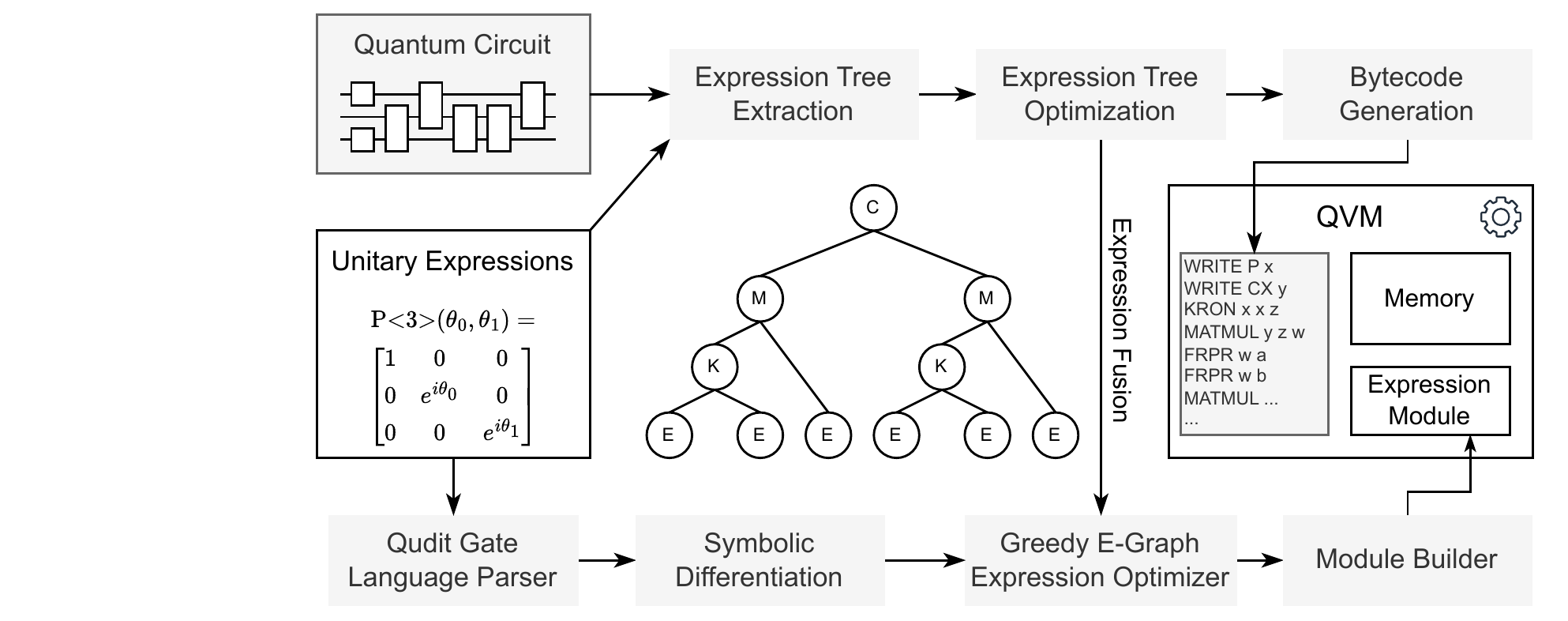}
    \caption{Overview of the two compilation
      pipelines in OpenQudit working towards efficient simulation of
      quantum circuits and their gradients. The bottom pipeline begins
      with unitary expressions formatted as Qudit Gate Language
      programs. It differentiates and optimizes these programs before
      exporting them as a module for later use by a Qudit Virtual
      Machine. The top pipeline, responsible for generating optimized
      instructions to simulate quantum circuits, first extracts an
      expression tree by solving the tensor contraction ordering
      problem over the input circuit. Further optimizations are
      performed on the resulting expression tree, including expression
      fusion, before final bytecode generation.}
    \label{fig:overview}
\end{figure}

We take a practical approach to extend quantum programming languages and systems to meet all the functionality requirements mentioned earlier. This work introduces the Qudit Gate Language (QGL), an embeddable unitary-specific expression language. Gates and other higher-order unitary-valued functions can be written in QGL as \textit{unitary expressions}, parsed, and verified for equality based on their mathematical expressions rather than their names. Additionally, to integrate QGL with optimizing quantum compilers, we provide the Qudit Virtual Machine (QVM), a library on top of QGL that accelerates and enhances quantum programming language optimizations and translations. QVM accomplishes this by accelerating unitary and gradient evaluations for circuits composed of QGL unitary expressions. This particular stage is absent from other symbolic systems, and we argue that it is mandatory for performance, extensibility, and interoperability. Finally, to enhance portability and cohesion between the components, we introduce a novel qudit circuit intermediate representation (QCIR) utilizing expression equality rather than label equality. Unlike other quantum IRs, unitary expressions are the basic instructional unit here. QCIR is designed to be flexible, efficient, and, most importantly, assist with the safe composition of tools by automatically detecting gate expression discrepancies.

Together, these components comprise the OpenQudit system. The library's primary goals are to lower the barriers to describing new gates, shift the responsibility of gate identification from labels to expressions, provide useful symbolic reasoning tools around expressions, and guide program optimization. These, in turn, will encourage and facilitate novel hardware architecture design and interoperability between programming systems.

Toward the same goal of accelerating quantum design and implementation, our language and system are not biased toward the binary form of quantum computing. Hence, Qudit is in all proposed names rather than Qubit. Qudits are units of quantum information larger in dimension than qubits, their binary counterparts. Qudits have been experimentally demonstrated in various quantum systems, including trapped ions~\cite{ringbauer2022universal}, photonic systems~\cite{malik2016multi,reck1994experimental}, neutral atoms~\cite{anderson2015accurate}, and superconducting devices~\cite{goss2022high,qutritrb}. They have greater computational capacity and, as a result, are garnering a growing interest. Algorithm designers have successfully applied qudits in areas such as simulation~\cite{gustafson2022noise}, optimization~\cite{bottrill2023exploring}, and cryptography~\cite{groblacher2006experimental}, among other fields. These applications demonstrate that quantum programs effectively utilizing qudits achieve asymptotic performance improvements~\cite{gokhale2019asymptotic}. Yet, software stacks and programming systems have largely neglected qudits.

The primary issue is a lack of literature and industry consensus on "standard" gates for qudits. Moreover, most problems require more gate definitions due to a larger dimension. It then becomes challenging for engineers to develop a general-purpose gate library traditionally. By natively supporting qudit unitary expressions in QGL, we can address this challenge and enable the organic growth of qudit gate libraries.


\textbf{Overview.} OpenQudit establishes two compilation pipelines: one for parsing and symbolic differentiation of QGL programs and another for optimizing, generating, and numerically processing code for quantum programs using QVM. Figure~\ref{fig:overview} illustrates the overall approach.

The basic QGL abstraction is the unitary expression, which symbolically describes a gate's functionality. For productivity, unitary expressions are succinctly stated as strings and parsed into the internal format rather than requiring users to program directly to the internal representation API. The system symbolically differentiates and optimizes QGL programs (unitary expressions) using e-graphs and equality saturation methods~\cite{egraph} employing the EGG~\cite{egg} library. The automatic symbolic differentiation functionality is highly desirable as gradients are complex to compute and optimize manually. Compared with other symbolic approaches, at this level, we introduce the notion of {\it symbolic circuit congruence}, which extends symbolic circuit equivalence to ignore the global phase in quantum programs and consider the existence of mappings between representations.

After a phase of symbolic manipulation, QGL programs are compiled into an LLVM~\cite{llvm} Module and passed to the QVM. The QVM can then reference these expressions and their gradients from a jitting LLVM context. The QVM has a larger goal of computing full circuit unitaries and gradients. This process proceeds by extracting an expression tree from a quantum circuit, optimizing the tree, and then generating QVM Bytecode. The QVM instructions describe the computation required for circuit transformations at the numerical optimization level: e.g. {\tt MATMUL} - matrix multiply, {\tt KRON} - Kronecker product, etc. Combining a memory buffer for intermediate computation, the bytecode to execute a circuit, and an expression module containing primitive methods, the QVM can efficiently calculate a circuit's unitary and gradients. In the most common cases, 3-qubit subcircuit evaluations, we have demonstrated slightly above a 10 times speedup against BQSKit, the state-of-the-art numerical quantum framework.

To summarize, the key contributions of this work are:
\begin{enumerate}
    \item The design and implementation of the Qudit Gate Language (QGL), an embeddable, qudit-capable, gate expression domain-specific language.
    \item QGL program just-in-time compilation pipeline including symbolic differentiation and optimization via a greedy simultaneous e-graph extraction.
    \item A scalable, qudit-capable, quantum circuit intermediary utilizing expression equality rather than label equality, supporting all possible gates by definition and leading to more robust translations between quantum programming languages, software development toolkits, and execution frameworks.
    \item A circuit simulation compilation pipeline, bytecode, and virtual machine optimized for repeatedly evaluating a quantum circuit's unitary and gradients.
    \item A detailed evaluation against state-of-the-art tools.
\end{enumerate}

\section{Background}
\label{sec:background}

\subsection{E-Graphs and Equality Saturation}
E-graphs are data structures designed to efficiently represent and reason about equivalences among expressions across various domains, particularly in symbolic computation, program optimization, and automated theorem proving. An e-graph consists of equivalence classes (e-classes), each containing multiple equivalent nodes (e-nodes). The e-nodes represent expressions, with directed edges pointing to e-classes that indicate their relationships. This structure allows for a compact representation of potentially infinite sets of equivalent expressions.

Equality saturation is a technique for automatically exploring a large set of equivalent expressions by systematically applying transformation rules. In this technique, a given e-graph is repeatedly transformed using a set of specified rules, with the goal of saturating the graph with all equivalent forms that can be reached through these rules. The graph is considered saturated when a round of rule applications creates no new nodes. This process results in sets of expressions deemed equivalent under the transformation rules. 

By leveraging e-graphs to store equivalences, equality saturation can be performed more efficiently than other techniques, as it avoids recomputing equivalences or generating redundant expressions. A saturated e-graph can lead to equivalent expression transformations by evaluating each e-node in a specific e-class. In particular, this approach is beneficial in compiler design, where discovered simplifications can enhance performance and robust equality checks can detect bugs.

\subsection{Quantum State}
The state of a binary quantum computer is maintained in quantum bits,
or \textit{qubits}. A qubit exists in some superposition of two basis
states. Most often, for computational applications, these basis states
are $\ket{0}$ and $\ket{1}$, represented by the vectors
$\left[ \begin{smallmatrix} 1 \\ 0 \end{smallmatrix} \right]$ and
$\left[ \begin{smallmatrix} 0 \\ 1 \end{smallmatrix} \right]$,
respectively. The superposition of a qubit $\ket{\phi}$ is then given
by the equation $\ket{\phi} = \alpha\ket{0} + \beta\ket{1} =
\left[ \begin{smallmatrix} \alpha \\ \beta \end{smallmatrix} \right]$
with complex coefficients $\alpha$, $\beta$. Pure quantum states --
ideal states in a perfectly closed system -- have coefficients,
referred to as \textit{amplitudes}, constrained such that $|\alpha|^2
+ |\beta|^2 = 1$. Higher-dimensional quantum states can be represented
by qu\textit{d}its, where \textit{d} refers to the qudit's
\textit{radix}, or the number of basis states in
superposition. Qutrits, ququarts, and ququints are commonly used to
refer to qudits with radix 3, 4, and 5, respectively. A pure qudit
state is defined similarly $\ket{\phi_{(d)}} =
\sum_{i=0}^{d-1}{\alpha_i\ket{i}}$ with $\alpha_i \in \mathbb{C}$
constrained by $\sum_{i=0}^{d-1}{|\alpha_i|^2} = 1$.

Composing quantum states is represented by the outer product of the
individual states, $\ket{\phi} = \ket{\phi_a} \otimes
\ket{\phi_b}$. For example, the state of a two-qubit system is given
by $\ket{\phi} = \ket{\phi_a} \otimes \ket{\phi_b} =
\left[ \begin{smallmatrix} \alpha_a\alpha_b \\ \alpha_a\beta_b
    \\ \beta_a\alpha_b \\ \beta_a\beta_b \end{smallmatrix} \right]$,
which is a superposition over the two-qubit basis states $\ket{00},
\ket{01}, \ket{10}, \ket{11}$. In this example, the composed state is
\textit{seperable} since there exist individual states $\ket{\phi_a}$
and $\ket{\phi_b}$, such that $\ket{\phi} = \ket{\phi_a} \otimes
\ket{\phi_b}$. However, not all composed states are separable. These
inseparable states are said to be \textit{entangled}. The notation $D$
refers to the dimension of an entire system, whereas previously $d$
referred to the dimension of a single qudit. It is not always the case
for a $n$-qudit system that $D = d^n$ since mixed-radix systems are
valid, although uncommon in practice. Formally $D =
\prod_{i=0}^{n-1}{d_i}$. For example, a system composed of a qubit and
a qutrit has $D=2\times3=6$ basis states.

\subsection{Quantum Operations}
A quantum state is transformed by matrix operators that maintain the
abovementioned pure quantum constraints. By definition, these linear transformations belong to the unitary group $U(D)$ with infinitely many valid operations. Additionally, operations can be either constant or parameterized by a finite number of real variables.

While any \textit{unitary} is a valid state transformation, each quantum hardware architecture provides a small, fixed set of natively executable operations due to engineering and physical constraints. Quantum computers that are \textit{universal} are capable of implementing any operation with a sequence of their native instructions.

\textit{Gates}, local operations affecting the state of only some qudits in a larger system, are described compactly by small
unitary operators. The term, gate, is often used interchangeably with quantum instruction, operation, or unitary. Some ubiquitous standard gates are given in
Figure~\ref{fig:gates}. Converting local gates into their global
operations involve mapping the small unitary and the specified qudit
indices into a full-system size matrix: $\text{extend}(G, I) = U$
where $G$ is a $k$-qudit gate, $I = (i_0, i_1, ... i_{k-1})$ is the
qudit indices the gate is applied to, and $U$ is the resulting global
unitary operator that applies $G$ on qudits specified by $I$. This
\textit{extension} process is done first through a series of outer
products with appropriately sized identity matrices and followed with
a matrix permutation~\cite{mike_and_ike}.

\begin{figure}
    \centering
    $$RX(\theta) = \begin{bmatrix}
        \cos{\frac{\theta}{2}} & -i\sin{\frac{\theta}{2}} \\
        -i\sin{\frac{\theta}{2}} & \cos{\frac{\theta}{2}}
    \end{bmatrix} RY(\theta) = \begin{bmatrix}
        \cos{\frac{\theta}{2}} & -\sin{\frac{\theta}{2}} \\
        -\sin{\frac{\theta}{2}} & \cos{\frac{\theta}{2}}
    \end{bmatrix} RZ(\theta) = \begin{bmatrix}
        e^{-i\frac{\theta}{2}} & 0 \\
        0 & e^{i\frac{\theta}{2}}
    \end{bmatrix}$$
    
    $$U1(\lambda) = \begin{bmatrix}
        1 & 0 \\
        0 & e^{i\lambda}
    \end{bmatrix} U2(\phi, \lambda) = \frac{1}{\sqrt{2}}\begin{bmatrix}
        1 & -e^{i\lambda} \\
        e^{i\phi} & e^{i(\phi + \lambda)}
    \end{bmatrix} U3(\theta, \phi, \lambda) = \begin{bmatrix}
        \cos{\frac{\theta}{2}} & -e^{i\lambda}\sin{\frac{\theta}{2}} \\
        e^{i\phi}\sin{\frac{\theta}{2}} & e^{i(\phi + \lambda)}\cos{\frac{\theta}{2}}
    \end{bmatrix}$$
    
    $$CNOT = \begin{bmatrix}
        1 & 0 & 0 & 0 \\
        0 & 1 & 0 & 0 \\
        0 & 0 & 0 & 1 \\
        0 & 0 & 1 & 0
    \end{bmatrix} RZZ(\theta) = \begin{bmatrix}
        e^{-i\frac{\theta}{2}} & 0 & 0 & 0 \\
        0 & e^{i\frac{\theta}{2}} & 0 & 0 \\
        0 & 0 & e^{i\frac{\theta}{2}} & 0 \\
        0 & 0 & 0 & e^{-i\frac{\theta}{2}}
    \end{bmatrix} Phase3(\theta_0, \theta_1) = \begin{bmatrix}
        1 & 0 & 0 \\
        0 & e^{i\theta_0} & 0 \\
        0 & 0 & e^{i\theta_1}
    \end{bmatrix}
    $$
    \caption{Several standard gate definitions. The first two rows give many examples of parameterized single-qubit operations. CNOT and RZZ are two-qubit entangling operations. While there is a common understanding of many qubit-based operations, there are more potential qutrit and higher-dimensional qudit gates with no similar standard set. The qutrit phase operator is an example of an executable native gate on many qutrit-based computers.}
    \label{fig:gates}
\end{figure}

\subsection{Quantum Circuits}
\label{sec:circuits}

\begin{figure}
    \centering
    \includegraphics[trim={1.8cm, 0.3cm, 1cm, 0.1cm},clip,width=\linewidth]{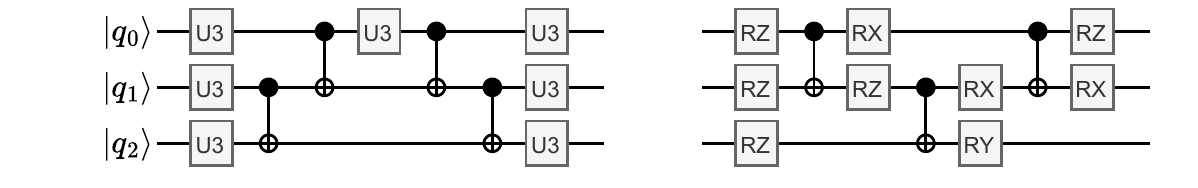}
    \caption{Two quantum programs illustrated in the circuit model. The left circuit comprises U3 and CNOT gates acting on three qubits. The starting state of the qubits is labeled on the left as $\ket{q_i}$; however, on the right, they are not labeled and are assumed to be indexed starting from the top and counting down. The right circuit uses a more diverse gate set that includes RX, RY, and RZ rotations.}
    \label{fig:circuit}
\end{figure}

Quantum \textit{programs} are typically expressed in the
\textit{circuit model} where wires extending from left to right
through time represent the qudits. Gates reflect operations applied to
the qudit's state when placed on the corresponding
wire. Figure~\ref{fig:circuit} illustrates the model. Unitary
operations capture all possible gates, but since the unitary group is
closed under multiplication, outer products, and permutations, the
semantics of purely quantum circuits can also be represented by
unitary operations. A circuit's unitary $C$ can be calculated by
taking a topologically ordered product of all extended gates:

$$C = \prod_{i}{\text{extend}(G_i, I_i)}$$

Circuits can also contain a small set of non-unitary
operations necessary for low-level operations, such as qudit state
reset, classically-controlled gates, and mid-circuit measurement. This
work does not focus on the semantics of programs containing these
instructions: the methods described in the later
sections can be applied to programs containing these instructions by
simply splitting the program around them. We also note that none of the existing symbolic manipulation infrastructures can reason directly about these non-unitary operations.

\section{Symbolic Circuits and Their Congruence}
\label{sec:symbolic}

Parameterized quantum circuits can represent an infinite set of
quantum programs, depending on the parameter values. We denote
parameterized circuits as $\symc{\cdot}$. These circuits are formally
defined as a map from their parameters to a unitary matrix, $\symc{C}:
\mathbb{R}^m \rightarrow U(D)$, where the circuit has $m$ parameters
and dimension $D$. Plugging in parameters, $\llbracket
C\rrbracket(\vec{p})$, results in a unitary operation describing the
resulting instantiated program's semantics.

Two quantum circuits are mathematically \textit{equivalent} if their respective unitary matrices are equal. This leads to a strict definition of equivalence for symbolic circuits, widely used in practice:
\begin{definition}[\textit{Symbolic Circuit Equivalence}]
Given two \textit{symbolic circuits}, $\symc{U}: \mathbb{R}^m \rightarrow U(D)$ and $\symc{V}: \mathbb{R}^m \rightarrow U(D)$, then
$$\symc{U} = \symc{V} \Longleftrightarrow \forall\vec{p}\in\mathbb{R}^m : \symc{U}(\vec{p})=\symc{V}(\vec{p}).$$
\end{definition}

On the other hand, from an operational or physical perspective,
this mathematical definition is unnecessarily strict, since two states
differing by a \textit{global phase} -- a complex factor lying on the
unit circle -- are indistinguishable due to quantum measurement. As a result, in most scenarios, quantum operations are interchangeable if they differ only by the global phase they install in output states.

\begin{definition}[\textit{Global Phase Congruence}]
Given two symbolic circuits, $\symc{U}: \mathbb{R}^m \rightarrow U(D)$ and $\symc{V}: \mathbb{R}^m \rightarrow U(D)$, then
$$\symc{U} \equiv_g \symc{V} \Longleftrightarrow \forall\vec{p}\in\mathbb{R}^m \ \ \exists \theta \in \mathbb{R} : \symc{U}(\vec{p})=e^{i\theta}\symc{V}(\vec{p}).$$
\end{definition}

Existing symbolic infrastructures use the notion of equivalence and
sometimes account for the global phase. Motivated by our practical
experiences, in OpenQudit, we introduce a novel notion of {\it symbolic
congruence}.

First, when working with experimentalists on studying and developing qudit gates, we noticed that errors often occur in practice. Multiple times, while composing software infrastructures, we transcribed their symbolic expressions with errors in the parameter expressions: we switched parameters, divided $\pi$ by a wrong constant factor, or forgot to negate an expression, among other simple mistakes. A more general circuit equality testing procedure helps in these usage cases by precluding hard-to-track bugs.

Second, such a procedure can improve compilation quality and speed. One observation is that in the context of program optimization, rewrite rules with simple permutations of their parameters can be easily applied, potentially leading to the discovery and application of powerful rewrites. Furthermore, division by 2 (or some other constant) is a common first step to most parameters; in Figure~\ref{fig:gates}, nearly all gates have all their parameters divided by 2. Removing this step leads to significant optimization in numerical compilers, which evaluate unitaries and gradients (hundreds of) thousands of times during compilation, depending on circuit size and gate count -- as a result, dividing by a constant accounts for many floating-point operations in numerical circuit evaluations driven by functional representations.

\begin{definition}[\textit{Symbolic Circuit Congruence}]
Given two symbolic circuits, $\symc{U}: \mathbb{R}^m \rightarrow U(D)$ and $\symc{V}: \mathbb{R}^m \rightarrow U(D)$, and a sequence of functions $f_0, f_1, ..., f_{m-1}$ with $f_i: \mathbb{R}^m \rightarrow \mathbb{R}$ $\forall i$, then
$$\symc{U} \equiv \symc{V} \Longleftrightarrow \forall\vec{p}\in\mathbb{R}^m \ \ \exists \theta \in \mathbb{R} : \symc{U}(\vec{p})=e^{i\theta}\symc{V}(f_0(\vec{p}), f_1(\vec{p}), ..., f_{m-1}(\vec{p})).$$
\end{definition}

Our extension idea is straightforward. Two circuits are congruent if we can find a mapping between their parameters—symbolic expressions within the unitary expression—which makes them symbolically equivalent to a global phase. Note that this captures the generic notion of physical circuit equivalence, but we use a different term to avoid confusion with the existing symbolic quantum circuit transformation literature.

\section{Qudit Gate Language}
\label{sec:qgl}

The Qudit Gate Language (QGL) describes qudit quantum gates as unitary expressions. The syntax is summarized in Extended Backus-Naur Form below with the unitary start symbol, and a few examples are shown in Figure~\ref{fig:qgl}.

\begin{lstlisting}[language=Python, basicstyle=\footnotesize\ttfamily, keywordstyle=\color{blue}, caption=Qudit Gate Language Grammar, label=grammar, captionpos=b]
unitary    ::= 'utry' ident [ radices ] '(' [ varlist ] ')' '{' expression '}' ;
radices    ::= '<' intlist '>' ;
expression ::= term { ('+' | '-') term } ;
term       ::= { '~' } factor { ('*' | '/') factor } ;
factor     ::= primary { '^' primary } ;
primary    ::= variable | constant | function | matrix | '(' expression ')' ;
matrix     ::= '[' row { ',' row } [ ',' ] ']' ;
row        ::= '[' exprlist ']' ;
exprlist   ::= expression { ',' expression } [ ',' ] ;
intlist    ::= integer { ',' integer } [ ',' ] ;
varlist    ::= variable { ',' variable } [ ',' ] ;
function   ::= ident '(' [ exprlist ] ')' ;
constant   ::= integer [ '.' integer ] ;
integer    ::= digit+
variable   ::= ident ;
ident      ::= letter { letter | digit } ;
digit      ::= '0' | '1' | '2' | '3' | '4' | '5' | '6' | '7' | '8' | '9' ;
letter     ::= 'a'..'z' | 'A'..'Z' | '_' | greek_letters ;
\end{lstlisting}

The unlisted ``greek\_letters'' production captures any UTF-8 capital
or lowercase Greek letter. Three variables, $i$, $e$, and $\pi$, are
reserved and imply their standard mathematical meanings. Therefore,
these variables cannot be declared in a unitary definition. Supported
functions include trigonometric functions $cos$, $sin$, $tan$, $sec$,
$csc$, $cot$, the natural logarithm $ln$, the exponential $exp$, the
power function $pow$, and the square root $sqrt$. Additionally, a QGL
program is semantically defined for any expression that can be
symbolically expressed element-wise in a closed form. This implies
that operations such as matrix multiplication and addition are
supported but not matrix exponentials. However, a matrix can be raised
to a constant integer power. Functions with matrix arguments are also
not supported. This set is powerful enough to express all gates described in literature.

Unitary definitions omitting a radix list are assumed to be qubit
only, requiring the dimension of the expression to be a power of
two. If omitted and the dimension is not a power of two, parsing fails.
If radices are specified, then the dimension of the
expression is required to be the product of the radices.

After parsing, we store each unitary definition in a 2D array of
complex expressions. Each complex expression is a data structure
containing a real and imaginary symbolic expression. For simplicity,
the final parsed symbolic expressions have all trigonometric functions
converted to sine and cosine. In this internal representation, we further support operations over the parsed expressions, such as symbolic matrix multiplication, kroneckor product, substitution, matrix embedding, conjugation, and transposition. These operations enable forming larger complex expressions from simple ones, such as controlled, inverse, or fused operations.

\begin{figure}[b]
\centering
\begin{minipage}{0.22\textwidth}
\begin{lstlisting}[language=C, basicstyle=\footnotesize\ttfamily, keywordstyle=\color{blue}]
utry CNOT() {
  [
    [1, 0, 0, 0],
    [0, 1, 0, 0],
    [0, 0, 0, 1],
    [0, 0, 1, 0],
  ]
}
\end{lstlisting}
\end{minipage}
\hfill
\begin{minipage}{0.35\textwidth}
\begin{lstlisting}[language=C, basicstyle=\footnotesize\ttfamily, keywordstyle=\color{blue}, mathescape=true]
utry U2($\phi$, $\lambda$) {
  [
      [1, ~e^(i*$\lambda$)],
      [e^(i*$\phi$), e^(i*($\phi$+$\lambda$))]
  ] / sqrt(2)
}
\end{lstlisting}
\end{minipage}
\hfill
\begin{minipage}{0.30\textwidth}
\begin{lstlisting}[language=C, basicstyle=\footnotesize\ttfamily, keywordstyle=\color{blue}, mathescape=true]
utry P<3>($\theta_0$, $\theta_1$) {
  [
    [1, 0, 0],
    [0, e^(i*$\theta_0$), 0],
    [0, 0, e^(i*$\theta_1$)],
  ]
}
\end{lstlisting}
\end{minipage}
\caption{Three examples of quantum gate unitary expressions in the Qudit Gate Language. The constant CNOT 2-qubit gate on the left requires no parameters. The U2 gate in the middle uses Greek letters and a matrix-scalar operation for readability. The qutrit phase gate on the right specifies the radices ``<3>'', implying a single-qutrit gate.}
\label{fig:qgl}
\end{figure}

\subsection{Expression Optimization}
OpenQudit utilizes the \textit{e-graph} data structure and \textit{equality saturation} methods~\cite{egraph} implemented in the \textit{EGG} library~\cite{egg} to optimize expressions and their derivatives. We extracted a foundational set of rewrite rules for equality saturation over real expressions from Herbie~\cite{herbie} and refined them with Enumo~\cite{enumo}. This refined set of rules proved sufficient to identify all closed-form trigonometric identities available on Wikipedia.

To perform simplification, we first initialize an {\it e-graph} with all symbolic real and imaginary components from both the unitary and gradient expressions. We then execute equality saturation with EGG using constant folding and otherwise the default settings. Finally, for each initial element in the {\it e-graph}, we extract an equivalent expression according to a custom cost function. The types of expressions and their associated costs are presented in Table~\ref{tab:eggcost} and were empirically determined to provide meaningful simplification over unitary expressions.

\begin{wraptable}{r}{0pt}
    \begin{tabular}{|c|c|}
        \hline
        Expression Type & Cost \\
        \hline
        $\pi$, Variable & 0.0 \\
         Constant & 0.5 \\
         $\mathtt{\sim}, +, -$ & 1.0 \\
         $*, /$ & 5.0 \\
         $\text{sqrt}, \sin, \cos$ & 50.0 \\
         $\exp, \ln, \text{pow}$ & 100.0 \\
        \hline
    \end{tabular}
    \caption{Expression Cost}
    \label{tab:eggcost}
\end{wraptable}

Optimal simultaneous extraction is typically performed with Integer
Linear Programming (ILP)~\cite{ilpextraction}, which can be
challenging to scale in practice. Instead, we adopt a greedy bottom-up
approach. After complete equality saturation, we repeatedly store and
update costs for each {\it e-class} in the graph until no change is
observed. The cost is calculated by calculating the score for each
{\it e-node} in the {\it e-class} and then picking the minimum {\it e-node}. If the
score for an {\it e-node} is incomplete -- because its children have not yet
been scored -- an infinity is assigned.

After stabilizing the cost of each {\it e-class}, we extract the first
expression. We select the {\it e-node} that minimizes each
{\it e-class} along the traversal down the graph. To encourage future
extractions to utilize common subexpressions, we set the cost of each
traversed {\it e-class} to zero. Our rationale is that during execution, this
expression will have been computed already and, ideally, reside in
registers. Following each extraction, we must update the cost of all
{\it e-classes} based on the prior bottom-up process, ensuring the new
zero-cost expressions propagate throughout the entire graph. Although
this extraction algorithm is not guaranteed optimal, it offers a
favorable trade-off between compilation time and expression quality.

\subsection{Equality and Congruence}
OpenQudit also uses equality saturation to determine equivalence,
global phase congruence, and conservatively symbolic
congruence. Mathematical equivalence is the most straightforward to
evaluate with equality saturation. To check if two expressions are
equivalent, start an {\it e-graph} with the two expressions and perform
equality saturation. If, at termination, they are in the same {\it e-class},
then they are equivalent. To check if two matrices are equal, we
simply check the equality between every element. If there is a hint
that the two matrices will be equal, we can accelerate the check by
building the {\it e-graph} with every element and only saturating once. This
will have the opposite effect if the expressions are unequal since the
full matrix saturation is more expensive, and the first equality check
on the elements usually fails.

Congruence up to a global phase introduces the additional challenge of
first identifying the phase. Assuming that the two expressions differ
only by a phase, every pair of elements will only differ by the same
global phase. We take the quotient of the first pair of non-zero
elements. If the quotient is $1$ or $-1$, the phase difference is $0$
or $\pi$. Otherwise, if a phase exists, the quotient will be
simplifiable to the form $\cos{\theta} + i\sin{\theta}$. We then use a
simplified guided equality saturation
approach~\cite{koehler2024guided} to extract a top-level sine
expression from the imaginary component. Focusing on the imaginary
component to extract the phase ensures the sign is always correct. If,
instead, we extracted a cosine from the real component, we would need
to additionally determine if the phase must be negated because
$\cos(\theta) = \cos(-\theta)$. Once the phase has been identified, we
multiply it into one of the matrices and then perform the mathematical
equivalence check. This ensures all the elements differ only by
the discovered phase.

We conjecture that proving symbolic circuit congruence is an
undecidable problem; however, we provide an algorithm that determines
congruence reasonably well for many quantum gates and gate
expressions. This algorithm fixes a left-hand side (LHS) and
right-hand side (RHS) expression, where the LHS is being evaluated
against. We start by simplifying the LHS by replacing common
expressions with fresh variables. For example, the U3 gate would have
$\frac{\theta}{2}$ replaced with $\theta^{'}$. Then, we enumerate a
parameter alphabet $\Sigma$ containing all the current variables of
LHS, as well as mathematical expressions involving them. In the case
of the U3, this may look like: $\{0, \frac{\pi}{2}, \pi, \theta^{'},
\phi, \lambda, \frac{\theta^{'}}{2}, \frac{\phi}{2},
\frac{\lambda}{2}, 2*\theta^{'}, \theta^{'}+\phi, \phi +
\frac{\pi}{2}...\}$. Then, the algorithm iterates over every subset of
$m$ elements from the alphabet, where $m$ is the number of variables
in RHS, and substitutes these for RHS's variables. With each
substitution, we employ the previous global phase congruence check,
and on success, we terminate returning the discovered congruence
relationship.

\section{Quantum Circuit Intermediate Representation}
\label{sec:ir}

This section introduces our novel quantum circuit intermediate representation (QCIR) designed explicitly for qudits. This representation utilizes expression-based identity checks for quantum gates, offering a robust and flexible framework for implementing complex quantum programs. Our IR is fully compatible with the Qudit Gate Language (QGL), facilitating seamless integration with all quantum gates expressible in QGL without requiring additional modifications beyond specifying the corresponding QGL code. Furthermore, QCIR supports control operations such as mid-circuit measurement, reset, and classically-controlled gates as first-class concepts. These features enable QCIR to represent complete quantum programs.

Internally, QCIR organizes quantum operations into cycles using a two-dimensional array abstraction. Each cycle stores a mapping from qudit and classical bit indices to instructions, which enables efficient manipulation and querying of the quantum circuit structure. The cycle structure represents a sequence of operations applied in parallel, allowing the intermediate to represent logical programs before compilation and physical circuits after compilation. Quantum circuit intermediates that store operations internally in a topologically sorted vector require conversions to other data structures, such as a DAG circuit representation, for specific stages of compilation. The cycle structure additionally contains the indices of the next and previous cycles for each instruction in the cycle, allowing one to iterate over a circuit as if it were a DAG.

To align with OpenQudit's overall mission of supporting quantum compilation frameworks, we also enable parameterized subcircuits to function as unit instructions. This feature allows compilers to utilize circuit partitioning explicitly, eliminating the need for additional systems and the corresponding conversions.

Each instance of the IR includes an indexed mutable gate set, allowing operations to refer to specific indices within the gate set. This design reduces memory usage for large circuits by caching gate objects specific to each circuit. The existence of a gate in the set is determined using equality testing based on QGL, ensuring that identity checks focus on expression identity rather than label identity. Combined with the previously mentioned equality and congruence checks, our intermediate representation is a valuable tool for translating between various quantum programming frameworks.
\section{Qudit Virtual Machine}
\label{sec:qvm}
Complementing the just-in-time compilation functionality of the qudit gate language is the qudit virtual machine (QVM) and its associated compilation pipeline. The QVM is designed to facilitate repetitive evaluations of unitaries and their gradients, as is common in numerical quantum compilation. It consists of three main components: the bytecode to be executed, an expression module providing context for expression evaluations, and pre-allocated matrix buffers for intermediate computation space. The bytecode results from the circuit compilation pipeline, starting with a parametric quantum circuit. The expression module is the JIT-compiled unitary and gradient subroutines for every expression involved in the circuit. Figure~\ref{fig:overview} illustrates the compilation pipelines and the resulting QVM.

\subsection{Expression Module}
\label{sec:module}
When a circuit is set to be compiled for the QVM, it begins with a scan for every unitary expression present. These expressions are then extracted and compiled into a QGL program, which is differentiated and optimized before being built into an executable module. This process involves lowering the optimized symbolic representation into an LLVM IR module. During execution, the Just-In-Time (JIT) compilation functionality of LLVM is utilized to provide highly optimized callable methods for each expression and its gradient.

The only optimization we perform during code generation is common subexpression elimination (CSE). We achieve this by maintaining a map of evaluated expressions to their corresponding registers. While the JIT engine within LLVM is capable of finding these optimizations, we perform CSE to ensure a smaller code size and to avoid unnecessary loads and stores. We also require that all unitary input buffers for QGL-generated subroutines be initialized to the identity matrix. For many gates, like the U1 and qutrit phase gate, this practice reduces the number of necessary writes since there are already ones along the diagonal. In most cases, this either improves performance or has no impact at all, as every element will need to be written regardless.

\subsection{Expression Trees}
\begin{figure}
    \centering
    \begin{subfigure}{0.73\linewidth}
    \includegraphics[trim={0.7cm 0.4cm 0.7cm 0.4cm},clip,width=\textwidth]{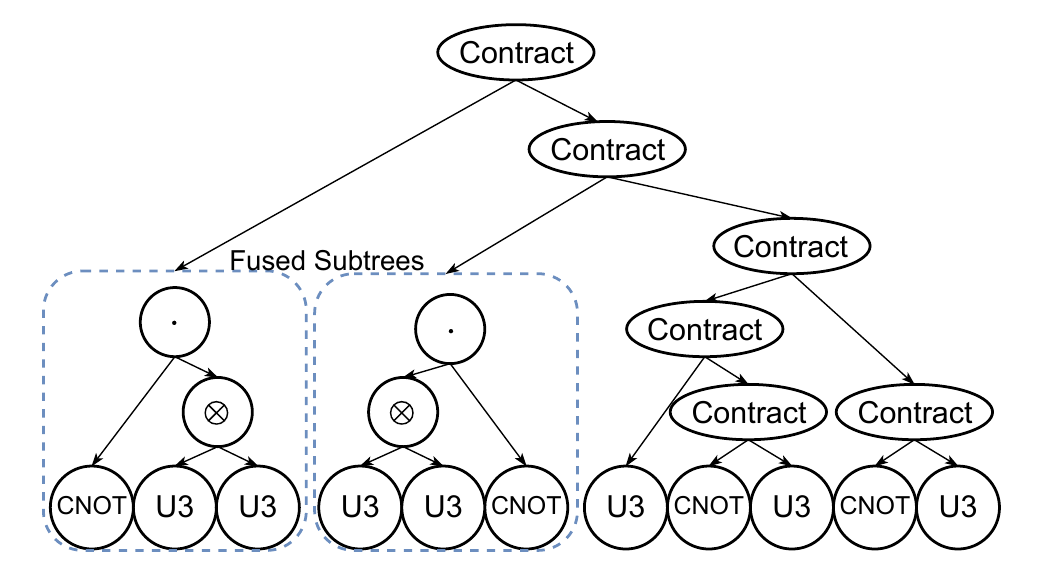}
    \end{subfigure}
    \hfill
    \begin{subfigure}{0.25\linewidth}
    \includegraphics[trim={0.72cm 1.305cm 0.75cm 1.61cm},clip,width=\textwidth]{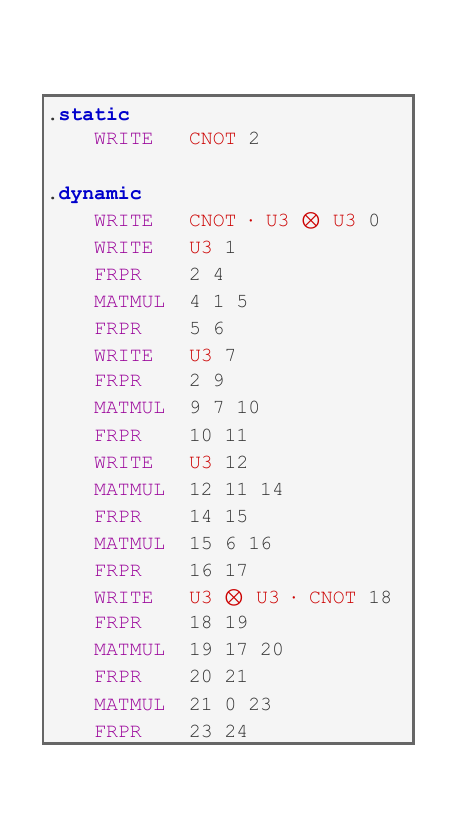}
    \end{subfigure}
    \caption{The expression tree intermediate and final bytecode resulting from compiling the example circuit in Figure~\ref{fig:circuit}. The compilation process fused two subtrees, requiring new fused expressions to be compiled in parallel with this simulation compilation. The resulting bytecode is split into two sections: static and dynamic. The static section is only executed once, whereas the dynamic section is executed every time, assuming the static code has already been run.}
    \label{fig:treeandcode}
\end{figure}

The most impactful step in the resulting QVM execution speed is parsing a quantum circuit into an expression tree. Here, the expression tree details the sequence of operations necessary to simulate the quantum circuit. Each leaf is a QGL unitary expression, and every non-leaf node is either a multiplication, kroneckor (outer) product, permutation, or contraction node. The root node outputs the result for the entire circuit. See Figure~\ref{fig:treeandcode} for an example expression tree. While there are many equivalent expression trees, all outputting the same, many will require vastly different amounts of computation. This is because performing more operations locally accomplishes more with smaller operations.

A quantum circuit can be translated to a tensor network, and as a result, parsing an expression tree is equivalent to solving for a tensor contraction ordering over a tensor network. Finding the optimal ordering is NP-hard in general~\cite{o2019parameterization} and is intractable in practice. For compilation time, we use a simple greedy approach~\cite{schindler2020algorithms} with one lookahead. The only difference from the standard greedy strategy we implement is performing an outer product when certain criteria are met. First, the pair of tensors must be less than two qudits each, and second, the result of the outer product must directly result in a matrix multiplication. This case arises commonly during synthesis-based compilation as a two-qudit gate followed by two single-qudit gates. In these small-sized cases, directly performing the outer product and matrix multiplication is more performant than tensor contraction.

\subsection{Tree Optimization}
In preparation for bytecode generation, we perform three optimizations on an expression tree. First, we detect and fuse subtrees. Then, we fuse contraction permutations, and finally we perform constant propagation.

Since every leaf node is an QGL program, we can symbolically evaluate parent nodes, JIT the fused expression, and replace the entire subtree with a new leaf. This must be done selectively as fusing too large of an operation can slow down both the compilation and execution times. Currently, QGL fusion only supports outer products and matrix multiplication. As a result, we select subtrees that are two qudits or less that contain only outer-products, matrix-multiplications, and unitary expressions. In Figure~\ref{fig:treeandcode}, you can see the fused subtrees and the resulting expressions in the bytecode.

Considering two gates as tensors and contracting over them is most efficiently done in four steps. First, the left tensor's indices are permuted and combined, then the right's indices. At this point, the tensors are now matrices and are multiplied, effectively summing over the contracted indices. Finally, the output matrix will have its indices separated and unpermuted. This process is equivalent to multiplying the extended matrices as in section~\ref{sec:circuits}, but will typically require much fewer operations. This speed-up is because the two matrices multiplied will not be $D\times D$; rather, their dimension will be determined by the number of indices that are summed. Suppose we are chaining many tensor contractions, as is common in quantum circuit simulation. In that case, we can fuse the output index permutation of a child contraction node with the input permutation of a parent contraction. This chaining technique will remove unnecessary data movement and speed up circuit calculations.

The last optimization we perform is constant propagation, which simply marks subtrees as constant or unparameterized. This is important during bytecode generation, as these constant subtrees will only be calculated once.

\subsection{Bytecode}
Since the QVM is optimized for repetitive executions, QVM bytecode is split into two sections, static and dynamic. The static section is entirely constant and only ever computed once on the first run. The dynamic section is parameterized and evaluated every time. Figure~\ref{fig:treeandcode} provides an example of the bytecode necessary to execute one of the circuits in Figure~\ref{fig:circuit}. There are four instructions in the QVM-compatible bytecode necessary to describe any quantum simulation:
\begin{itemize}
    \item The $\mathtt{WRITE}$ operation evaluates a unitary expression by calling its associated JIT-compiled function in the expression module, storing the result in a specified matrix buffer. This operation is the only one with an operand that is not a matrix buffer as it requires a function pointer into the expression module.
    \item The $\mathtt{FRPR}$ or fused-reshape-permute-reshape command is an accelerated subroutine for pre- and post-processing matrices for tensor contraction. This operation is equivalent to reshaping the input matrix to a tensor, permuting the indices, and then reshaping back to a matrix. The specifics of this operation are defined by the shape of the input and output matrices, the dimensions of the intermediate tensor, and the tensor index permutation. By fusing these sub-operations into one, we can avoid needing to support tensor object types in the implementation, as the input and output are both matrices.
    \item The $\mathtt{MATMUL}$ operation is matrix multiplication between two matrix buffers with a separate buffer for output storage.
    \item The $\mathtt{KRON}$ is the outer product or kroneckor product of two matrices with a separate buffer for output storage.
\end{itemize}

Execution of the bytecode starts with an initial warm-up phase. In this phase, the unitary and gradient computation buffers are first zero-allocated. Then, the unitary buffers are initialized to the identity matrix to support the optimization described in section~\ref{sec:module}. Finally, the QVM executes the static code section. At this point, the QVM has officially started and can execute dynamic code. If the QVM is set to compute gradients, each buffer index refers to a vector of buffers, and the instructions perform the unitary evaluation and differentiation automatically.

\section{Evaluation}
\label{sec:evaluation}
We implemented OpenQudit with about 15,000 lines of Rust code utilizing the faer~\cite{faer} mathematical library rather than standard BLAS libraries. This provided a substantial speed-up on small matrix multiplications commonly performed as part of the system. For generating and JITting the expression modules, we used LLVM~17~\cite{llvm} with the aggressive (-O3) optimization level. OpenQudit will be made publicly available on acceptance; it is currently private for the double-blind review process.

As OpenQudit provides many features, we performed several experiments to compare each with state-of-the-art tools. Depending on the test, we compared with Qiskit 1.2.4, pytket 1.34.0, BQSKit 1.2.0, and JAX 0.4.35. JAX is configured for CPU usage only and is programmed to calculate unitaries with the extension method in section~\ref{sec:background}. All evaluations were conducted on an AMD EPYC 7702 processor with 1TB of memory. All tools were set to default settings, leading to all except Qiskit using one core, whereas Qiskit used all cores for the simulation.

\subsection{QGL Evaluation}
\begin{figure}[b]
    \centering
    \includegraphics[width=\linewidth]{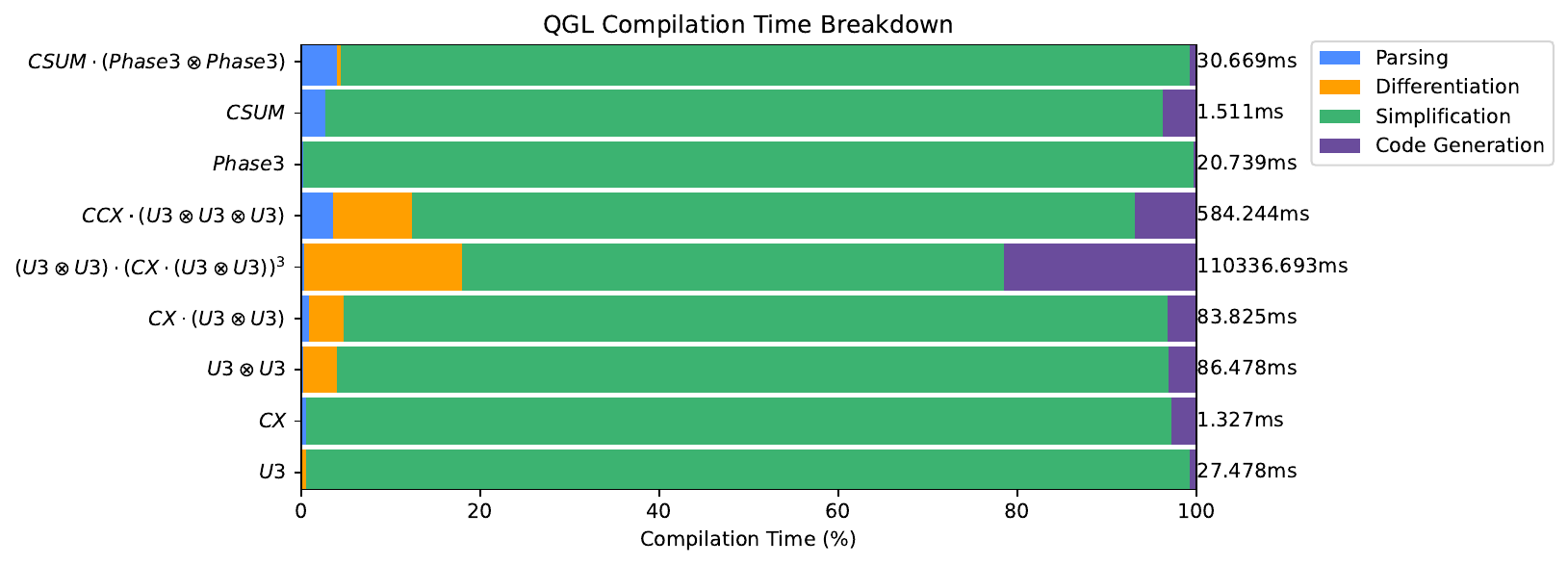}
    \caption{Breakdown of the time spent in each stage of QGL compilation across many common expressions.}
    \label{fig:jitbreakdown}
\end{figure}

First, we evaluate the QGL compilation pipeline. In this experiment, we ask two questions:
\begin{enumerate}
    \item How does the performance of QGL compiled code compare with state-of-the-art JIT tools and quantum compiler frameworks?
    \item What is the cost and scaling of compilation?
\end{enumerate}
These questions aim to determine if JITting QGL code can be effectively used in numerical quantum optimization. We selected a variety of common qubit, qutrit, and many-qubit block expressions appearing in many synthesis algorithms and compiled them with OpenQudit. Figure~\ref{fig:jitbreakdown} details the time for each compilation stage and the total time. The e-graph-based simplification stage consumes most of the compilation time. Simplification is broken into two steps: equality saturation and expression extraction. After further profiling this stage, we found that extraction consumed most of the time in every scenario. Most compilations require well under a second, however $(U3\otimes U3)\cdot(CX\cdot (U3\otimes U3))^3$ stands as an outlier requiring ~110 seconds for compilation. This expression can represent any two-qubit unitary and has 24 unique parameters, while the next largest has 9 unique parameters. Even though the cost of QGL compilation can be amortized during large-circuit compilation and there is room for optimization in the implementation, this example illustrates the problematic scaling of QGL for large expressions and the need for the accompanying QVM compilation pipeline to work over larger expressions.

\begin{table}[t]
\centering
\begin{tabular}{|c|c|c|c|c|c|}
\hline
\textbf{Expression} & \textbf{OpenQudit32} & \textbf{OpenQudit64} & \textbf{BQSKit} & \textbf{JAX} & \textbf{Qiskit} \\
\hline
$U3$ & 0.028 & 0.063 & 1.617 & 11.192 & 368.568 \\ 
$CX$ & 0.001 & 0.001 & 1.693 & 4.258 & 362.468 \\ 
$U3\otimes U3$ & 0.086 & 0.141 & 3.106 & 17.063 & 398.576 \\ 
$CX\cdot (U3\otimes U3)$ & 0.086 & 0.142 & 3.816 & 18.158 & 433.558 \\ 
$(U3\otimes U3)\cdot(CX\cdot (U3\otimes U3))^3$ & 0.588 & 0.953 & 11.415 & 61.296 & 592.898 \\ 
$CCX\cdot (U3\otimes U3\otimes U3)$ & 0.194 & 0.263 & 9.508 & 68.288 & 463.634 \\ 
$Phase3$ & 0.012 & 0.030 & 112.732 & 8.792 & * \\ 
$CSUM$ & 0.003 & 0.003 & 2.356 & 4.283 & * \\ 
$CSUM\cdot (Phase3\otimes Phase3)$ & 0.035 & 0.066 & 226.903 & 37.497 & * \\ 
\hline
\end{tabular}
\caption{Time taken to calculate an expression's unitary in microseconds. Results marked with * are not available due to lack of qudit support. The two different columns for OpenQudit reflect different floating point bit widths. BQSKit and Qiskit compute 64-bit floating point arithmetic, and JAX computes 32-bit floating point.}
\label{tab:utryexpr}
\begin{tabular}{|c|c|c|c|}
\hline
\textbf{Expression} & \textbf{OpenQudit32} & \textbf{OpenQudit64} & \textbf{BQSKit} \\
\hline
$U3$ & 0.037 & 0.074 & 5.310 \\ 
$CX$ & 0.001 & 0.001 & 3.966 \\ 
$U3\otimes U3$ & 0.143 & 0.210 & 13.396 \\ 
$CX\cdot (U3\otimes U3)$ & 0.162 & 0.232 & 16.524 \\ 
$(U3\otimes U3)\cdot(CX\cdot (U3\otimes U3))^3$ & 5.950 & 7.303 & 60.422 \\ 
$CCX\cdot (U3\otimes U3\otimes U3)$ & 1.106 & 1.381 & 49.650 \\ 
$Phase3$ & 0.014 & 0.032 & 125.714 \\ 
$CSUM$ & 0.003 & 0.003 & 5.689 \\ 
$CSUM\cdot (Phase3\otimes Phase3)$ & 0.068 & 0.174 & 265.121 \\ 
\hline
\end{tabular}
\caption{Time to calculate an expression's unitary and gradient in microseconds. The two different columns for OpenQudit reflect different floating point bit widths. BQSKit computes 64-bit floating point arithmetic.}
\label{tab:gradexpr}
\end{table}

Next, we execute the compiled code to evaluate the unitary and gradient compute times. Table~\ref{tab:utryexpr} compares the unitary compute time of OpenQudit with BQSKit, JAX, and Qiskit. OpenQudit is generic over floating-point precision, so we list two columns, one for 32-bit and another for 64-bit precision. In all cases, OpenQudit computed every expression's unitary in under a microsecond, while no other tool could compute even one under a microsecond. Qiskit does not support qudits, so there are blanks in the qutrit benchmarks. Qiskit spawns many threads for computation and, as a result, has a high constant cost. On average, OpenQudit32 computed the unitary 1.66, 2062.48, 972.81, and 64784.16 times faster than OpenQudit64, BQSKit, JAX, and Qiskit, respectively. Table~\ref{tab:gradexpr} presents the gradient evaluation times. Here, we do not compare against Qiskit and JAX since they do not support matrix gradients. OpenQudit32 computed the gradient 1.58 and 2126.18 times faster than OpenQudit64 and BQSKit, respectively.

While compiling QGL code can be done ahead of time and amortized over many numerical optimizations, it is still worth determining whether a single numerical optimization can justify compiling QGL code. We focus on the U3 gate, the most optimized and commonly used in the BQSKit framework. We evaluate how many gradient evaluations would be necessary to justify the compilation time. Each gradient evaluation of a U3 saves $5.310-0.037=5.273$ microseconds. This implies $27478/5.273=~5211$ computations are necessary to justify the compilation. Considering that even a tiny 3-4 qubit circuit may contain hundreds of U3 gates, which will need to be computed hundreds to thousands of times during gradient descent, it becomes clear that one numerical optimization is worth the compilation cost.

\begin{figure}
    \centering
    \includegraphics[width=\linewidth]{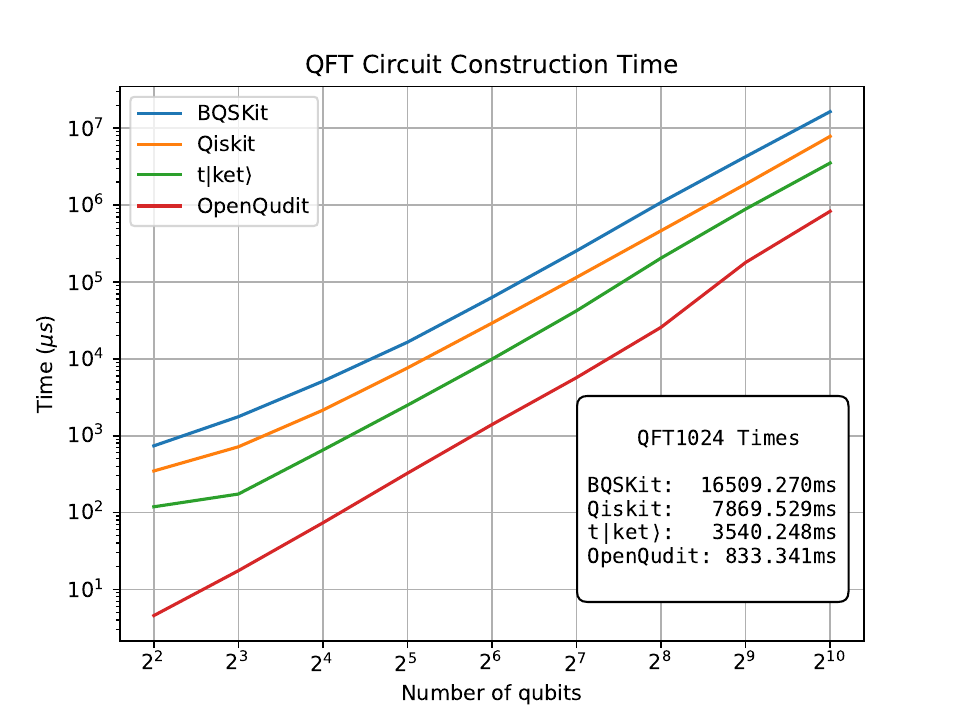}
    \caption{Construction time for various sizes of the Quantum Fourier Transform (QFT) circuit across OpenQudit and many popular frameworks.}
    \label{fig:qft}
\end{figure}

\subsection{Circuit Construction}
IBM stressed the importance of circuit manipulation performance in quantum compilation frameworks~\cite{nation2024benchmarking}. The reason is that circuits are loaded, split, and rebuilt during circuit compilation, among other operations. Slow performance in this area can affect the overall compilation performance. We construct a Quantum Fourier Transformation circuit~\cite{coppersmith1994ibm} with varying sizes in multiple frameworks to evaluate our quantum circuit intermediate representation. Figure~\ref{fig:qft} plots the results. OpenQudit constructed the circuit faster than all other frameworks and was the only one to build a 1024-qubit QFT in under one second. Across all sizes, OpenQudit built the circuit on average 9.36, 27.67, and 62.09 times faster than Tket, Qiskit, and BQSKit.

\subsection{QVM Evaluation}
\begin{figure}
    \centering
    \includegraphics[trim={0, 0, 1.2cm, 0.4cm},clip,width=\linewidth]{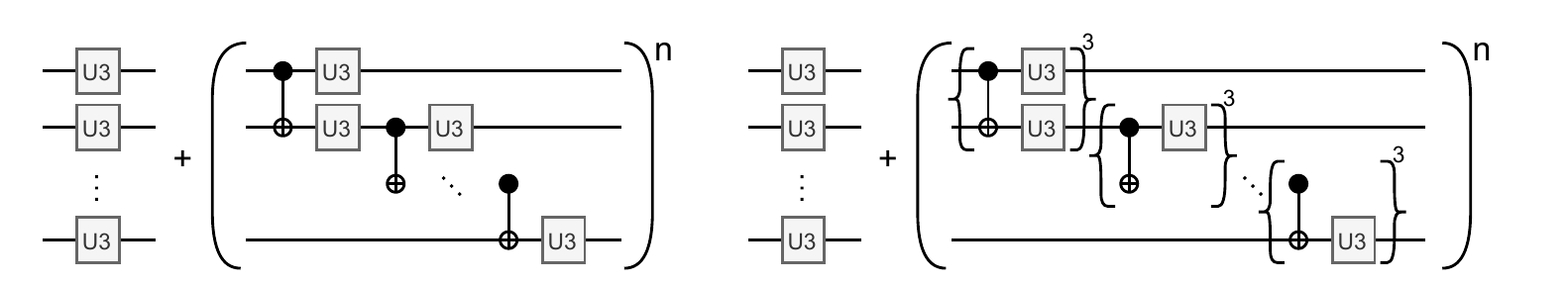}
    \caption{The structure of the square brickwall circuits benchmarked in the QVM evaluation. On the left is the thin variant, and on the right is the thick variant. Both variants start with U3 gates on all qubits and are followed by $n$ layers of gates where $n$ is the number of qubits. In the thin variant, each layer consists of a cnot followed by two U3s on every pair of consecutive qubits, forming a ladder structure. The thick variant is very similar but repeats the gate block 3 times on each pair. These circuits resemble circuits seen during numerical synthesis and span the difficulty spectrum as the thick variant allows for maximal local computation and the thin for very little local computation.}
    \label{fig:qsearch-circuit}
\end{figure}

To evaluate the QVM's performance, we constructed two types of square brick wall circuits ranging in size from 3 to 7 qubits. These circuits are built out of blocks of a CNOT followed by two U3 gates. They start with a U3 gate on each qubit, followed by a layer containing a block on every pair of consecutive qubits. This layer is repeated as many times as there are qubits in the circuit. The thin variant only contains one CNOT and U3 block per pair in each layer, whereas the thick variant contains three blocks per pair. Figure~\ref{fig:qsearch-circuit} illustrates the circuit structures. The thin variant allows for very little local computation before growing the size of operations to the circuit size. On the other side of the spectrum, the thick variant allows for maximal local computation, as three CNOTs on a pair of qudits is expressive over every two-qubit unitary. These structures resemble those found in typical state-of-the-art bottom-up synthesis algorithms, such as QSearch~\cite{qsearch}. We believe this makes the unitary and gradient evaluations performed on these circuits good proxies for how QVM may speed synthesis algorithms and, as a result, synthesis-based compilers.

\begin{figure}
    \centering
    \begin{subfigure}{0.49\linewidth}
        \includegraphics[trim={0.6cm 0.1cm 1.3cm 0.8cm},clip,width=\textwidth]{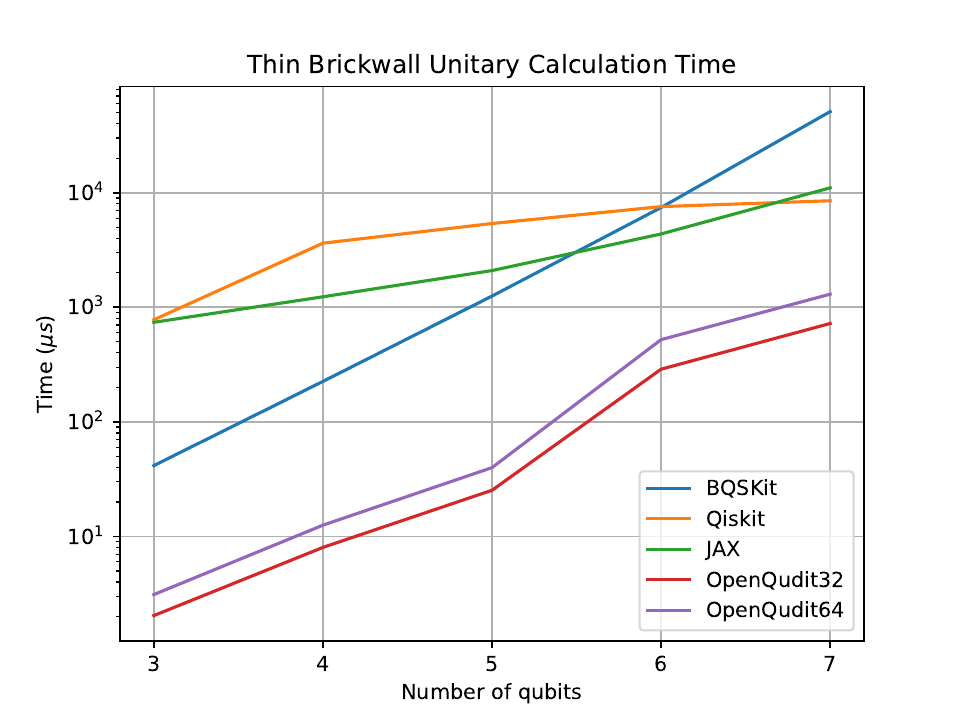}
    \end{subfigure}
    \hfill
    \begin{subfigure}{0.49\linewidth}
        \includegraphics[trim={0.6cm 0.1cm 1.3cm 0.8cm},clip,width=\textwidth]{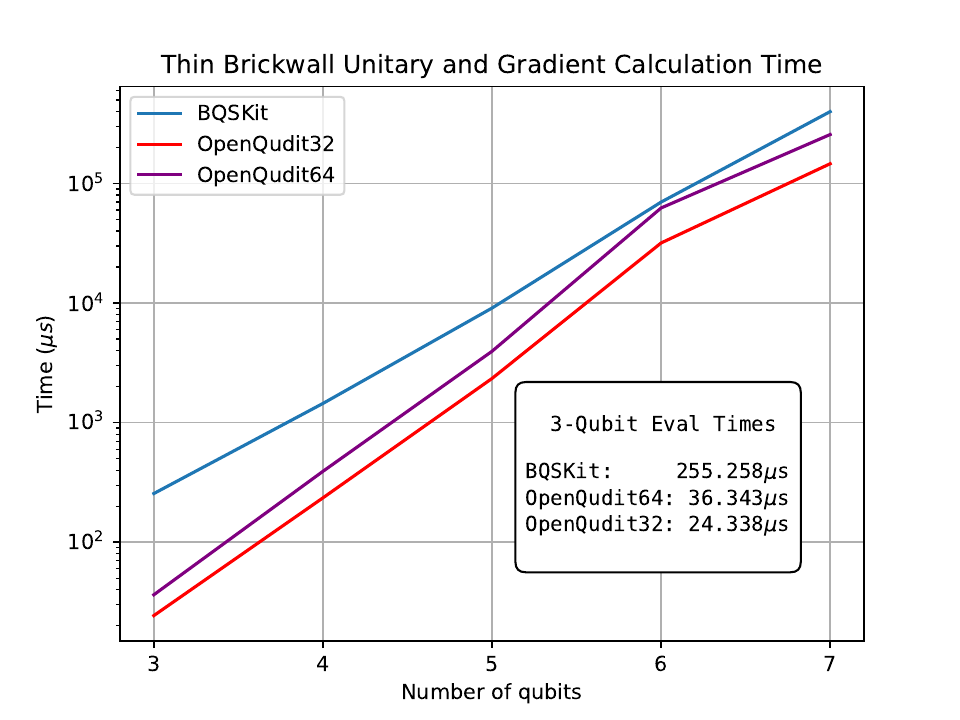}
    \end{subfigure}
    \begin{subfigure}{0.49\linewidth}
        \includegraphics[trim={0.6cm 0cm 1.3cm 0.8cm},clip,width=\textwidth]{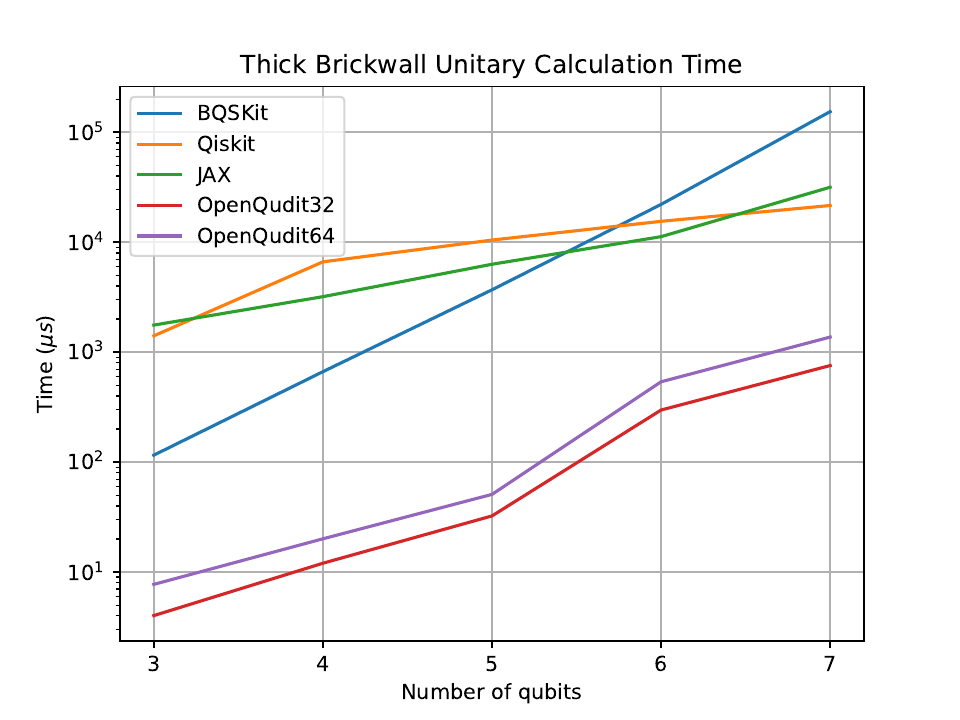}
    \end{subfigure}
    \hfill
    \begin{subfigure}{0.49\linewidth}
        \includegraphics[trim={0.6cm 0cm 1.3cm 0.8cm},clip,width=\textwidth]{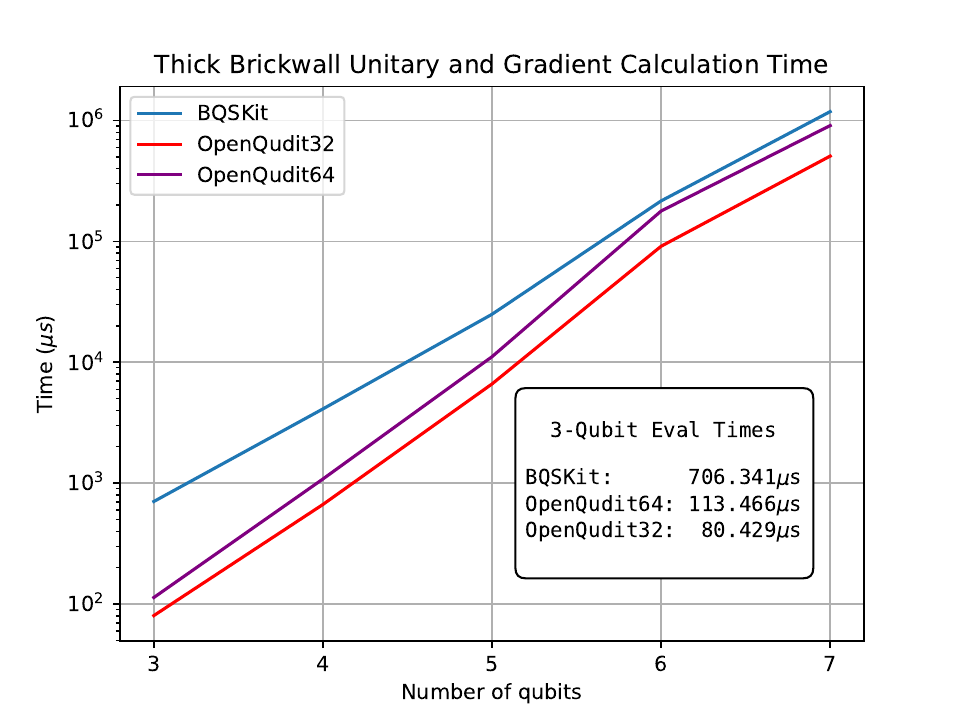}
    \end{subfigure}
    \caption{Evaluation times for thin and thick brickwall circuits. On the left, the unitary computation time is plotted; on the right, both the unitary and gradients are evaluated in each benchmark.}
    \label{fig:circuittime}
\end{figure}

Figure~\ref{fig:circuittime} plots the unitary and gradient calculation times. OpenQudit outperforms all other frameworks across the circuit sizes tested. OpenQudit32 evaluated circuits on average 1.66, 38.91, 126.05, and 217.09 times faster than OpenQudit64, BQSKit, JAX, and Qiskit, respectively. OpenQudit32 evaluated gradients on average 1.71 and 5.09 times faster than OpenQudit64 and BQSKit. However, OpenQudit32 evaluated the 3-qubit circuit gradients 10.49 times faster than BQSKit. There is no significant difference between the thin and thick benchmarks results, as all frameworks tested could efficiently compute the local operations. 

An exciting observation is that JAX outperformed BQSKit in larger sizes. BQSKit computes the unitary and gradient evaluations using a tensor-based approach, which should be asymptotically faster than the extension method described in Section~\ref{sec:circuits}. However, JAX discovered performance improvements even when programmed using the extension method. This shows the efficacy of JIT-compiled expressions, as JAX jit compiles these evaluations.

Although, the scaling of the OpenQudit gradient evaluations approaches BQSKit towards the high end of the range. We believe that this is due to two reasons. First, the matrix multiplication method we use does not perform cache blocking. Our implementation is well-optimized for small matrices but scales poorly. Second, our tensor contraction ordering is far from optimal. With small circuits, there is little variance between the worst and best orderings, which implies that our method finds good enough contraction orderings. However, with larger circuits, the greedy approach becomes far from optimal, leading to performance similar to other methods.

\section{Future Work}
\label{sec:discussion}
Several improvements can be made to the implementation of OpenQudit and its associated compilation pipelines. One of the most important enhancements would be a better algorithm for tensor contraction ordering, which could significantly improve overall performance and, more specifically, the scalability of the QVM. For instance, in the case of the 4-qubit thin brickwall circuit, the difference in unitary evaluation speed between the optimal contraction ordering and the greedy approach is about fivefold. Although determining the optimal contraction ordering on the fly may be intractable, there is considerable room for improvement.

Additionally, we should enable fusing over tensor contractions. Currently, fusing expressions are only supported for matrix multiplications and Kronecker products. By implementing expression fusing for tensor contractions, we could significantly increase the number of subtrees that QGL can optimize. However, this would require a more effective algorithm for identifying candidate subtrees to prevent the entire tree from being fused.

Lastly, we should explore multithreading and GPU support for the QVM. Currently, all computations are executed on a single thread on the CPU. For larger calculations, integrating parallel matrix multiplication algorithms designed for better scalability will improve efficiency. Furthermore, the four bytecode instructions can be easily mapped to the GPU, and we hypothesize that a well-designed system could achieve substantial speed improvements in this area as well.

\section{Conclusion}
\label{sec:conclusion}
In this work, we introduced the OpenQudit system and its core components: the QGL, a language that can express any gate as a unitary expression; the QCIR, an intermediate representation that utilizes expression-identity instead of label-identity and can be expressed over any QGL expression; and the QVM, a machine designed to evaluate the unitary of any circuit and its gradients quickly. Our vision is for scientists and engineers to utilize the entire system or its components to enhance existing and emerging quantum programming frameworks, addressing the five goals outlined in the introduction.

Implementing support for QGL provides expressibility, extensibility, and accessibility because it can express any gate and is easily programmable for practitioners. The use of expression-identity improves the safety and robustness of tools, particularly in composition. The added safety comes from identifying and reporting discrepancies between gates that share the same names but may have different implementations. For instance, TKet and Qiskit, two of the most widely used quantum frameworks, define the RX Gate, one of the most common gates, slightly differently. TKet multiples the input by pi before performing trigonometric operations. Such discrepancies can create bugs when converting between tools, which can often be necessary when executing on their associated hardware. Our system can identify these inconsistencies and potentially suggest the correct mapping through our symbolic congruence algorithm.

Qudit gates are less documented, and as a result, there are more discrepancies. We envision our system employed as a translation layer to assist with robust interoperability. System engineers may adopt the QCIR fully or support QGL in some form, enabling translations into QCIR. Once converted to QCIR, users can freely translate between all supporting systems with the aforementioned correctness checks. Additionally, QCIR provides a scalable circuit data structure alongside the QVM to provide efficient and flexible compilation, further adding to the scalability and accessibility introduced by our system. We conclude that with proper tooling such as those proposed here, gates in the form of unitary expressions provide an essential abstraction for quantum programming.

\begin{acks}
    I appreciate and am grateful for the valuable discussions with Costin Iancu, Wim Lavrijsen, Bert De Jong, and Koushik Sen that refined the ideas presented in this work.

This work was supported by the U.S. Department of Energy, Office of Science, Office of Advanced Scientific Computing Research under Contract No. DE-AC05-00OR22725 through the Accelerated Research in Quantum Computing Program MACH-Q project.
\end{acks}

\bibliographystyle{ACM-Reference-Format}
\bibliography{refs}

\end{document}